\documentclass[pra,twocolumn,preprintnumbers,amsmath,amssymb,superscriptaddress,showpacs,longbibliography]{revtex4-2}
\usepackage{tabularx,amsmath,amsfonts,amssymb,boxedminipage,graphicx,mathtools,appendix,physics}
\usepackage{epstopdf, float}
\usepackage{algorithm2e}
\usepackage{subfigure}
\usepackage[final]{hyperref}
\usepackage{csquotes}
\MakeOuterQuote{"}
\RestyleAlgo{boxruled}
\hypersetup{
	colorlinks=true,       
	linkcolor=blue,          
	citecolor=blue,        
	filecolor=magenta,      
	urlcolor=blue         
}

\begin{document}

\title{A Non-Convex Optimization Strategy for Computing Convex-Roof Entanglement}
\author{Jimmie Adriazola}
\affiliation{Department of Mathematics, Southern Methodist University, Dallas, Texas, USA}
\author{Katarzyna Roszak}
\affiliation{FZU - Institute of Physics of the Czech Academy of Sciences, 182 00 Prague, Czech Republic}

\begin{abstract}
We develop a numerical methodology for the computation of entanglement measures for mixed quantum states. Using the well-known Schr\"odinger-HJW theorem, the computation of convex roof entanglement measures is reframed as a search for unitary matrices; a nonconvex optimization problem. To address this non-convexity, we modify a genetic algorithm, known in the literature as differential evolution, constraining the search space to unitary matrices by using a QR factorization. We then refine results using a quasi-Newton method. We benchmark our method on simple test problems and, as an application, compute entanglement between a system and its environment over time for pure dephasing evolutions. We also study the temperature dependence of Gibbs state entanglement for a class of block-diagonal Hamiltonians to
provide a complementary test scenario with a set of entangled states that are qualitatively different.
We find that the method works well enough to reliably reproduce entanglement curves, even for comparatively large systems. To our knowledge, the modified genetic algorithm represents the first derivative-free and non-convex computational method that broadly applies to the computation of convex roof entanglement measures.
\end{abstract}

\date{\today}
\maketitle

\section{Introduction}
Quantification of entanglement for \textit{pure} quantum states is reasonably straightforward because classical correlations cannot be present in fully quantum states. Thus, any measured correlation is quantum, and this implies that entanglement between two parts of a pure quantum system is inversely proportional
to the purity of the state of one (either) part of the system that is obtained after tracing out
the other. Measures such as entanglement entropy \cite{bennett96b} operate in this fashion, that is, they quantify entanglement by quantifying the purity of these subsystems (via von Neumann entropy here, but any measure of purity is usable, such as linear entropy \cite{santos00}, etc.).

Unfortunately, realistic quantum systems are rarely pure, due to outside influences such as
interactions with uncontrollable (often quantum) environments \cite{touzard19,katz23}, classical noise sources \cite{connors19}, and difficulties related to reliable repetition of experiments. The presence of noise
leads to the need for density matrices instead of quantum
state vectors, since density matrices are capable of capturing statistical properties of quantum ensembles. Such states can contain classical correlations (and quantum-classical \cite{modi14}), making the detection of entanglement through the purity of part of the system alone impossible. 

The naive solution to this is to quantify the entanglement of the mixed state by
calculating the average of the entanglement of the pure state of the eigendecomposition of the density matrix. This fails because of the way the set of \textit{separable} states is defined:
Any state is separable if and only if it is possible to decompose it into
a mixture of pure separable states. 
The definition has a deeper physical meaning that quantum correlations cannot be obtained
via local operations and classical communication (LOCC) \cite{nielsen99,bennett99}.

A trivial example would be an equal mixture of two maximally entangled Bell states,
\begin{equation}
\hat{\rho}_{\mathrm{sep1}}=\frac{1}{2}|\Psi_+\rangle\langle \Psi_+|+\frac{1}{2}|\Psi_-\rangle\langle \Psi_-|,
\end{equation}
with $|\Psi_{\pm}\rangle=(|01\rangle\pm|10\rangle)/\sqrt{2}$.
Since each Bell state $|\Psi_{\pm}\rangle$ is maximally entangled, the state $\hat{\rho}$ should also be
maximally entangled, but it is in fact separable. This is evident once it is written in the separable basis
$\{|ij\rangle\}$, with $i,j=0,1$, where we get 
\begin{equation}
\hat{\rho}_{\mathrm{sep1}}=\frac{1}{2}|01\rangle\langle 01|+\frac{1}{2}|10\rangle\langle 10|.
\end{equation}
The state is correlated, but only classically.
In less trivial cases, the decomposition that fulfills the definition and demonstrates separability
cannot be obtained via a change of basis (since it is not an eigendecomposition) and is typically much
harder to find. Such mixed states also include other mixtures of Bell states, such as separable X-states
\cite{mazurek14b},
e.~g.~$\hat{\rho}_{\mathrm{sep2}}=1/4 |00\rangle\langle 00|+1/4|11\rangle\langle +1/4
|\Psi_+\rangle\langle \Psi_+|$
or separable Werner states \cite{hiroshima00}. Prescriptions to find explicit separable decompositions
for two-qubit states can be found in Ref.~\cite{wootters98}, while Werner states are discussed
in Ref.~\cite{azuma06}.

To overcome this, the notion of convex roof entanglement measures was developed. Convex-roof entanglement measures are based on the idea that
the actual amount of entanglement
can be quantified as an average of pure-state entanglement of some pure-state decomposition
of a state, but the pure-state decomposition has to be the one that minimizes entanglement.
If the von Neumann entropy is used as the purity measure in the quantification of pure state
entanglement then the convex roof entanglement measure is called Entanglement of Formation
(EoF) \cite{bennett96,bennett96a}.
The minimization problem is complex because it requires the use of an in principle
infinite set of unitary matrices and
the number of minimization parameters grows with system
size $N$ as $N^2-1$. 

A side effect of the complexity of minimization is the relatively small number
of studies of mixed-state entanglement for larger systems. For two-qubit entanglement,
there exists a direct method for the quantification of the EoF \cite{wootters98} and consequently
there is an immense number of papers that discuss two-qubit correlations,
providing a detailed study of its, often curious, properties. For larger systems,
the dominant tool is Negativity \cite{plenio05b}, which does not relate to any measure of pure state entanglement and cannot quantify certain types of entanglement \cite{horodecki98}, but is much more accessible
numerically. 

Beyond that, studies of the EoF or other convex-roof entanglement measures 
are restricted to scenarios, where certain limitations on the symmetry of the density matrices 
are imposed \cite{audenaert01,wolf04,shiokawa09,jedrzejewski22}, which limit the
scope of the minimization. This does not mean that larger entangled states are just a generalization of two-qubit states. On the contrary, there exist properties characteristic
for qudits that cannot be reproduced or explained by qubit systems \cite{horodecki98,roszak18}
and not all of them can be captured by Negativity \cite{horodecki98}. However, the numerical resources required for the study of entanglement between larger systems are prohibitive
for systematic studies to be performed. 

An excellent methodology, due to R\"othlisberger et al~\cite{rothlisberger2009numerical} already exists for the efficient calculation of entanglement measures. This work even led to the development of a software library titled libCreme~\cite{rothlisberger2012libcreme}. In short, R\"othlisberger, and others, adapt gradient-based line searches on matrix manifolds~\cite{absil2008optimization} to compute convex roof entanglement measures. Although this work successfully paved the way for quantifying entanglement using powerful numerical optimization methods, the methodology suffers from two obvious flaws. First, deterministic line search methods are inefficient for identifying globally optimal points due to the curse of non-convexity. Indeed, despite its name, the computation of convex roof entanglement measures is not formulated as a convex optimization problem. Second, the computation of gradients on matrix manifolds is not automatic and impedes the adoptability of these gradient-based methods to a wide class of problems encountered in practice.

In this paper, we introduce a new algorithm for the efficient computation of convex roof entanglement measures. In principle, it can be treated as a black-box, and thus can be broadly applied for the computation of entanglement. Moreover, it is based on a genetic algorithm, a heuristic search method that addresses the non-convexity of the optimization problem. The method is then accelerated toward the nearest local minimum via a black-box implementation of a quasi-Newton method. Our intention with this hybrid, nonconvex-line search approach is to present
an efficient algorithm while maintaining a high level of usability for communities interested in quantifying entanglement in applications and less interested in the mathematical or computational details under the hood.

We test the algorithm on a series of examples for which the amount of entanglement
is known, or can be at least estimated. We choose examples that involve a strong
dependence on a parameter, such as time or temperature, so that
the level of success of the algorithm can also be evaluated on its capability to describe
the trends in entanglement behavior. This is particularly important in the study of entanglement
in realistic scenarios and in order to identify the properties of entanglement between
qudits that are not manifested in qubit-qubit entanglement. 

We use two-qubit examples to tune the parameters of the procedure for optimal operation and to demonstrate that the algorithm is capable of demonstrating sudden death of entanglement.
We then use it to quantify the evolution of entanglement between a qubit and an environment of many qubits.
We get smooth, reliable curves, which are expected in a unitary evolution as long as the overall purity of the
system is not too small. The last example pertains to the amount of entanglement found in a bipartite
temperature-equilibrium state. We find that at low temperatures, when the overall purity of the state is still large,
the algorithm performs well, and the obtained curves are smooth, but at larger temperatures the noisiness of the
results becomes non-negligible, although the overall trend for the temperature dependence of entanglement
is clearly visible.

The paper is organized as follows. In Sec.~\ref{sec2} we specify the optimization problem
under study. The modified genetic algorithm/quasi-Newton method that is used to tackle the optimization is 
discussed in Sec.~{\ref{sec3}.} Sec.~\ref{sec4} contains a number of examples of the
operation of the algorithm on examples with known or partially quantifiable entanglement. 
Sec.~\ref{sec5} concludes the paper. 

\section{The Optimization Problem \label{sec2}}
The computation of entanglement measures we pursue uses the well-known convex-roof construction~\cite{bennett96,aolita08}. To begin with its description, we define the notion of a pure state decomposition. Given a density matrix $\rho$ of rank $r$ acting on a Hilbert space $\mathcal{H}$ of finite dimension $d$, a pure state decomposition is defined by a set of $s\geq r$  probabilities $p_i$ and vectors $\left|\psi_i\right\rangle\in \mathcal{H}$ such that
\begin{equation}\label{eq:PSD}
\rho=\sum_{i=1}^s p_i\left|\psi_i\right\rangle\left\langle\psi_i\right|.
\end{equation}
The set of all pure-state decompositions is then simply given by
\begin{widetext}
\begin{equation*}
\mathfrak{D}(\rho)=\left\{\left\{p_i,\left|\psi_i\right\rangle\right\}_{i=1}^s, \left\{\left|\psi_i\right\rangle\right\}_{i=1}^s \subset \mathcal{H},\ p_i \geq 0, \ \sum_{i=1}^s p_i=1  \bigg| \rho=\sum_{i=1}^s p_i\left|\psi_i\right\rangle\left\langle\psi_i\right|\right\}.
\end{equation*}
\end{widetext}

The construction of convex roofs further requires a monotone entanglement $m:\mathcal{H}\to\mathbb{R}^+$ that quantifies the amount of entanglement in the system~\cite{vidal00}. The function $m$ is monotonically decreasing, under local operations and classical communication~\cite{chitambar2014everything}, and evaluates to zero if the density matrix $\rho$ is separable, i.e., $\rho$ can be written as a convex combination of product states. 

As a relevant example of an entanglement monotone, consider the entropy of entanglement. For a pure bipartite quantum state of a pure composite system $\rho_{AB}$, the  entropy of entanglement is given by the von Neumann entropy of the reduced density
matrix,
\begin{equation}\label{eq:EoE}
m_{\rm vN}\left(\rho_{AB}\right)=-\operatorname{Tr}\left[\rho_A \log \rho_A\right]=-\operatorname{Tr}\left[\rho_B \log \rho_B\right],
\end{equation}
where $\rho_A=\operatorname{Tr}_B\left(\rho_{A B}\right)$ and $\rho_B=\operatorname{Tr}_A\left(\rho_{A B}\right)$ are the reduced density matrices given as partial traces for each partition.

The convex-roof entanglement measure 
$M:\mathcal{H}\to\mathbb{R}^+$ is thus defined by the following optimization problem
\begin{equation}\label{eq:HardOpt}
M(\rho)=\inf _{\left\{p_i,\left|\psi_i\right\rangle\right\} \in \mathfrak{D}(\rho)} \sum_i p_i m\left(\left|\psi_i\right\rangle\right).
\end{equation}
In words, this computation constitutes a search for the smallest mean evaluation of a given entanglement monotone $m$ over the space $\mathfrak{D}$ of all pure state decompositions corresponding to a given density matrix $\rho$. As written, this search is intractable.

Fortunately, a theorem due to Schr\"odinger~\cite{schrodinger1935discussion} and Hughston, et al.~\cite{hughston1993complete}, colloquially known as the Schr\"odinger-HJW theorem, furnishes a unitary parameterization of the space $\mathfrak{D}$. More precisely, let $\mathcal{U}(k, r)$ denote the set of all $k \times r$ matrices $U \in$ $\mathbb{C}^{k \times r}$ with the property that $U^{\dagger} U$ equals the $r\times r$ identity matrix denoted by $\mathbb{I}_{r \times r}$. The theorem states that every $U \in \mathcal{U}(k, r)$ produces a pure-state decomposition $\left\{p_i,\left|\psi_i\right\rangle\right\}_{i=1}^k \in \mathfrak{D}(\rho)$ of the density matrix $\rho$. 

To see this, let $\lambda_i$ and $\left|v_i\right\rangle$ denote the eigenvalues and corresponding normalized eigenvectors of $\rho$ in the representation
\begin{equation}\label{eq:specdecomp}
\rho=\sum_{j=1}^r \lambda_j\left|v_j\right\rangle\left\langle v_j\right|.
\end{equation}
Note that since $\rho$ is a density matrix, it is by definition a positive definite matrix with a well-defined spectral decomposition. 
Now, given a matrix $U \in \mathcal{U}(k, r)$, define the auxiliary state
\begin{equation*}
\left|\varphi_i\right\rangle=\sum_{j=1}^r U_{i j} \sqrt{\lambda_j}\left|v_j\right\rangle, \quad i=1, \ldots, k,
\end{equation*}
corresponding to the spectral content of $\rho$ in Equation~\eqref{eq:specdecomp}. Observe that the auxiliary states $\left|\varphi_i\right\rangle$ yield a pure state decomposition through Equation~\eqref{eq:PSD} with the probabilities and state vectors given by
\begin{equation*}
p_i=\left\langle\varphi_i \big| \varphi_i\right\rangle, \qquad \left|\psi_i\right\rangle=\frac{1}{ \sqrt{p_i}}\left|\varphi_i\right\rangle.
\end{equation*}
Moreover, and this is key, the theorem states that every pure-state decomposition of $\rho$ can be realized from a matrix $U \in \mathcal{U}(k, r).$ Thus, the optimization problem can be reformulated as the matrix optimization problem
\begin{equation}\label{eq:EasierOpt}
\min _{k \geq r} \inf _{U \in \mathcal{U}(k, r)}  J(U)=\min _{k \geq r} \inf _{U \in \mathcal{U}(k, r)}  \sum_{i=1}^k p_i (U)m\left(\left|\psi_i(U)\right\rangle\right),
\end{equation}
where it is implied that $p_i$ and $\psi_i$ are computed from the Schr\"odinger-HJW method applied to a given density matrix $\rho.$
Although the Schr\"odinger-HJW theorem allows us to reformulate this problem as a more tractable search over unitary matrices, the ambiguity in the parameter $k$ forces us to perform a brute force search over the size of the unitary matrices furnishing the pure-state decompositions. We address all computational issues concerning this optimization problem in the following section.

\section{The Optimization Method\label{sec3}}

To search for the optimal unitary matrix in problem~\eqref{eq:EasierOpt}, we modify a preexisting optimization method due to Storn and Price~\cite{Storn,DiffEvoBook}.  The algorithm is called differential evolution (DE) and is a stochastic optimization method used to search for candidate solutions to generally non-convex optimization problems. The idea behind DE is inspired by evolutionary genetics and is thus part of a class of so-called genetic algorithms. Genetic algorithms are used in a wide context of optimization and design problems where derivative information of the objective function is inaccessible or difficult to compute.

DE searches the space of candidate solutions by initializing a population set of vectors, known as agents, within some region of the search space. These agents are then mutated (see Algorithm~\ref{algo:mut}) into a new population set, or generation. The mutation operates via two mechanisms: a weighted combination and a random "crossover." 

At each generation,  Algorithm~\ref{algo:mut}  generates a candidate $z$ to replace each agent $y$. In the mutation step, it chooses at random three agents $a$, $b$, and $c$ to create a new trial agent $\tilde{z}$ through the linear combination
\begin{equation*}
    \tilde{z}=a+F\cdot(b-c),
\end{equation*}
where $F\in[0,2]$. In the crossover step, a new candidate vector $z$ is constructed by randomly choosing some components $\tilde{z}$ and others from an additional randomly chosen agent $d$.
If we evaluate an objective functional $J:\mathbb{R}^N\to\mathbb{R}$ and find that $J(z)<J(y)$, then $z$ replaces $y$ in the next generation.

DE ensures that the objective functional $J$ of the optimization problem decreases monotonically with (the optimal member of) each generation. As each iteration "evolves" into the next, inferior agents "inherit" optimal traits from superior agents via mutation, or else they are discarded. After a sufficient number of iterations, the best vector in the final generation is chosen as the candidate global optimizer. A more detailed discussion about further implementation and benchmarking details can be found in the book by Price et al.~\cite{DiffEvoBook}.

As discussed, the classical DE algorithm is broken into two parts; a mutation step and an evolution step. The mutation step remains unchanged. However, the evolution step is modified so that candidate optimizers remain in the space $\mathcal{U}.$ To enforce the unitary constraint, we use the QR decomposition~\cite{trefethen2022numerical} which we now briefly review.

It is a fact that any real square matrix $A$ may be decomposed as
\begin{equation*}
A=Q R,
\end{equation*}
where $Q$ is an orthogonal matrix and $R$ is an upper triangular matrix. If instead $A$ is a complex square matrix, then $Q$ is unitary. More generally, we can factor a complex $k \times r$ matrix $A$, with $k \geq r$, as the product of a $k \times k$ unitary matrix $Q$ and a $k \times r$ upper triangular matrix $R$.

The key steps to adapting the DE method to the calculation of convex roof entanglement measures are as follows. To make DE mutation compatible with an optimization over unitary matrices, we must first reshape candidate optimizing matrices from $k\times r$ to $k*r\times 1$ since mutation is performed at the vector level. We then perform mutation via Algorithm~\ref{algo:mut}. The resulting mutated vector is then reshaped back into a $k\times r$ matrix $\tilde{Q}$ that is most likely no longer unitary because mutation does not preserve unitarity. 

In this step, we use the QR decomposition to ensure that a function evaluation of $J$ appearing in Equation~\eqref{eq:EasierOpt} is done so with a unitary matrix. That is, we apply a QR factorization in $\tilde{Q}$ to obtain a unitary matrix $Q$ of size $k\times k$ and delete its last $k-r$ columns so that it is a unitary matrix $U$ of size $k\times r$. The candidate matrix $U$ now belongs to the feasible optimization space $\mathcal{U}(k,r)$ as desired. This evolutionary process is summarized entirely by Algorithm~\ref{algo:HDE}.

Typically, DE calculates the convex roof entanglement measure correctly to one or two digits of accuracy. To refine the results and accelerate convergence, we use the DE output and feed it into a BFGS method~\cite{nocedal1999numerical} as implemented by MATLAB's \texttt{fminunc} function. Of course, inside of the evaluation of the entanglement measure, we perform a QR decomposition to ensure that the BFGS method remains constrained to searches over unitary spaces. The computation of the QR factorization is handled via MATLAB's built-in \texttt{qr} function.

One final comment we make is about the use of the QR decomposition to constrain our search space to unitary matrices. Perhaps, a more natural way to project onto this space is to find the closest unitary matrix $\hat{U}$ to $A$, in the sense of the operator norm. This would be facilitated by finding $\hat{U}$ such that
$$
\hat{U}=\arg \min _{\tilde{U}}\|A-\tilde{U}\|.
$$
The solution to this problem is well known~\cite{fan1955some}. The unitary matrix $\hat{U}$ that minimizes $\|A-\hat{U}\|$ can be calculated via the singular value decomposition (SVD):
$$
\begin{gathered}
A=\hat{V} \hat{\Sigma} \hat{W}^{\dagger}, \\
\hat{U}=\hat{V} \hat{W}^{\dagger}.
\end{gathered}
$$

In all computational examples, we consistently found that our results were more optimal when we adopted a QR approach over the SVD approach. We discuss further implementation details in the following section alongside our computational examples. 

 \begin{algorithm}[htbp]
\caption{Differential Mutation}\label{algo:mut}
\KwResult{A vector $z$ mutated from agents in a given generation as required by the DE Algorithm~\eqref{algo:HDE}.}
\SetKwInOut{Input}{Input}
 \Input{4 distinct members $a,b,c,d$ from the current generation of agents each with $N$ components, the crossover ratio $C_R\in(0,1),$ and weight $F\in(0,2)$.}
 \For{j=1:N}{
              Compute a random variable $\mathtt{rand}$\;
                 \eIf {$\mathtt{rand}<C_R$} {
                     $z[j]\gets a[j]+F*(b[j]-c[j])$
                     }
                     {$z[j]\gets d[j]$
                 }
         }
\end{algorithm}

 \begin{algorithm}[htbp]
\caption{Unitary Differential Evolution}\label{algo:HDE}
\KwResult{A vector likely to be globally optimal with respect to an objective $J$.}
\SetKwInOut{Input}{Input}
 \Input{A maximum number of iterations $\mathtt{Nmax}$, crossover ratio $C_R\in(0,1)$, weight $F\in(0,2)$, and matrix of size $k\times r$}
 Generate a random population of $k\times r$ unitary matrices \\
Reshape into a population \texttt{pop} of $N_{\rm pop}$ vectors of dimension $k*r$.\\
\While{$\mathtt{counter}<\mathtt{Nmax}$}{
 \For{$i=1:N_{\rm pop}$}{
 $\mathtt{CurrentMember}\gets \mathtt{Pop}_i$\;
     Choose three distinct vectors $a_i,b_i,c_i$ different from the vector $\mathtt{Pop}_i$\;
        Mutate $a_i,b_i,c_i,$ and the $\mathtt{CurrentMember}$ into the mutated vector $z$
        using the mutation parameters $C_R,F$ and Algorithm~\ref{algo:mut}\;
        Reshape $z$ and $\mathtt{CurrentMember}$ into $k\times r$ matrices $\tilde{Q}$ and $\tilde{U}$\;
        Perform a QR decomposition on the matrix $\tilde{Q}$ and assign the resulting $k\times k$ unitary matrix to $Q$\;
        Delete the last $k-r$ columns of $Q$ and assign this to ${U}$\;
         \If{$J({U})<J(\tilde{U})$} {
            Reshape ${U}$ into a $k*r$ dimensional vector $\tilde{z}$\;
             $\mathtt{TemporaryPop}_i\gets \tilde{z}$\;
         }
 }
 $\mathtt{Pop}\gets \mathtt{TemporaryPop}$\;
 $\mathtt{counter}\gets \mathtt{counter}+1$\;
}
 \end{algorithm}

\begin{figure*}[htbp]
\begin{centering}
\subfigure{\includegraphics[width=0.45\textwidth]{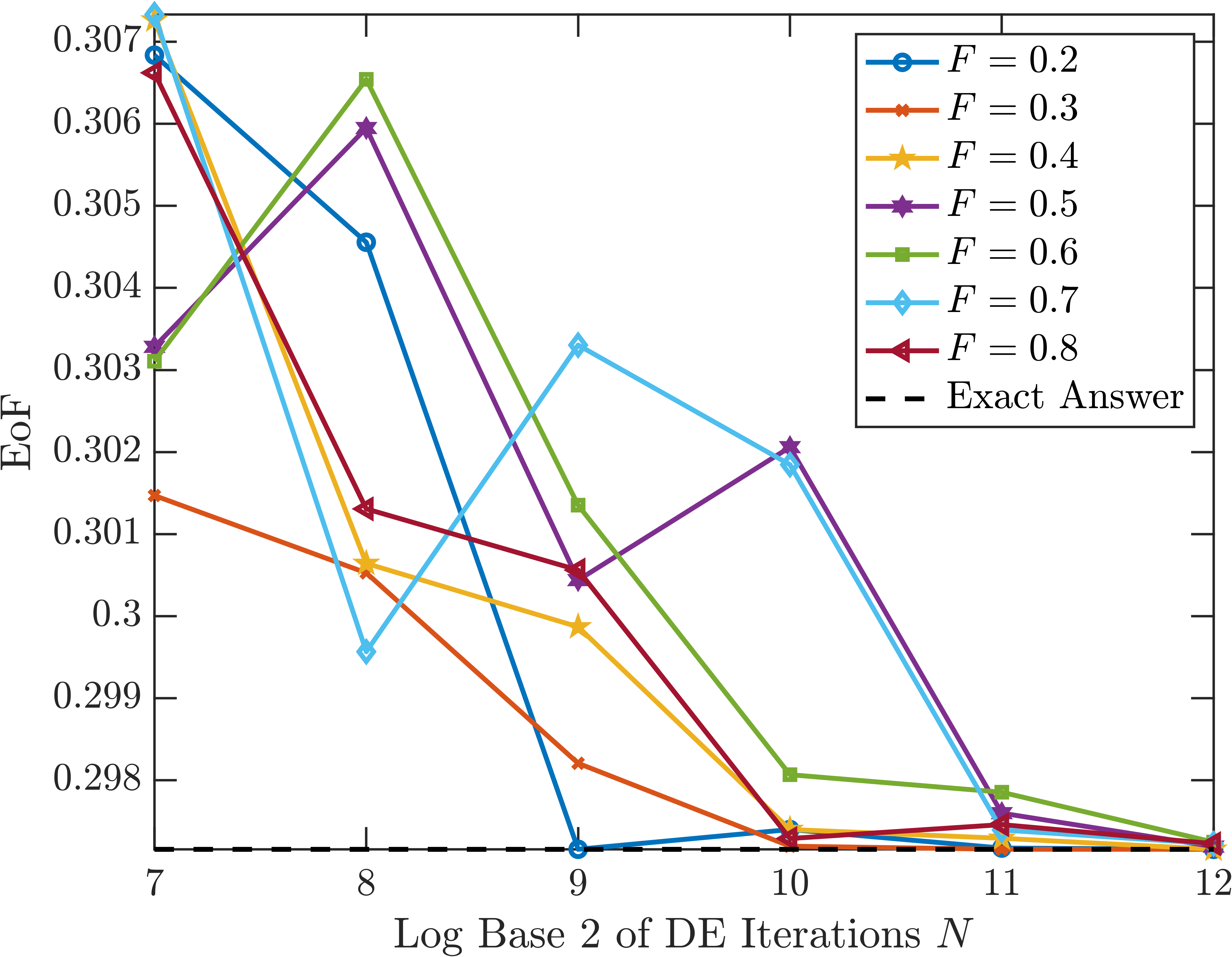}}
\subfigure{\includegraphics[width=0.45\textwidth]{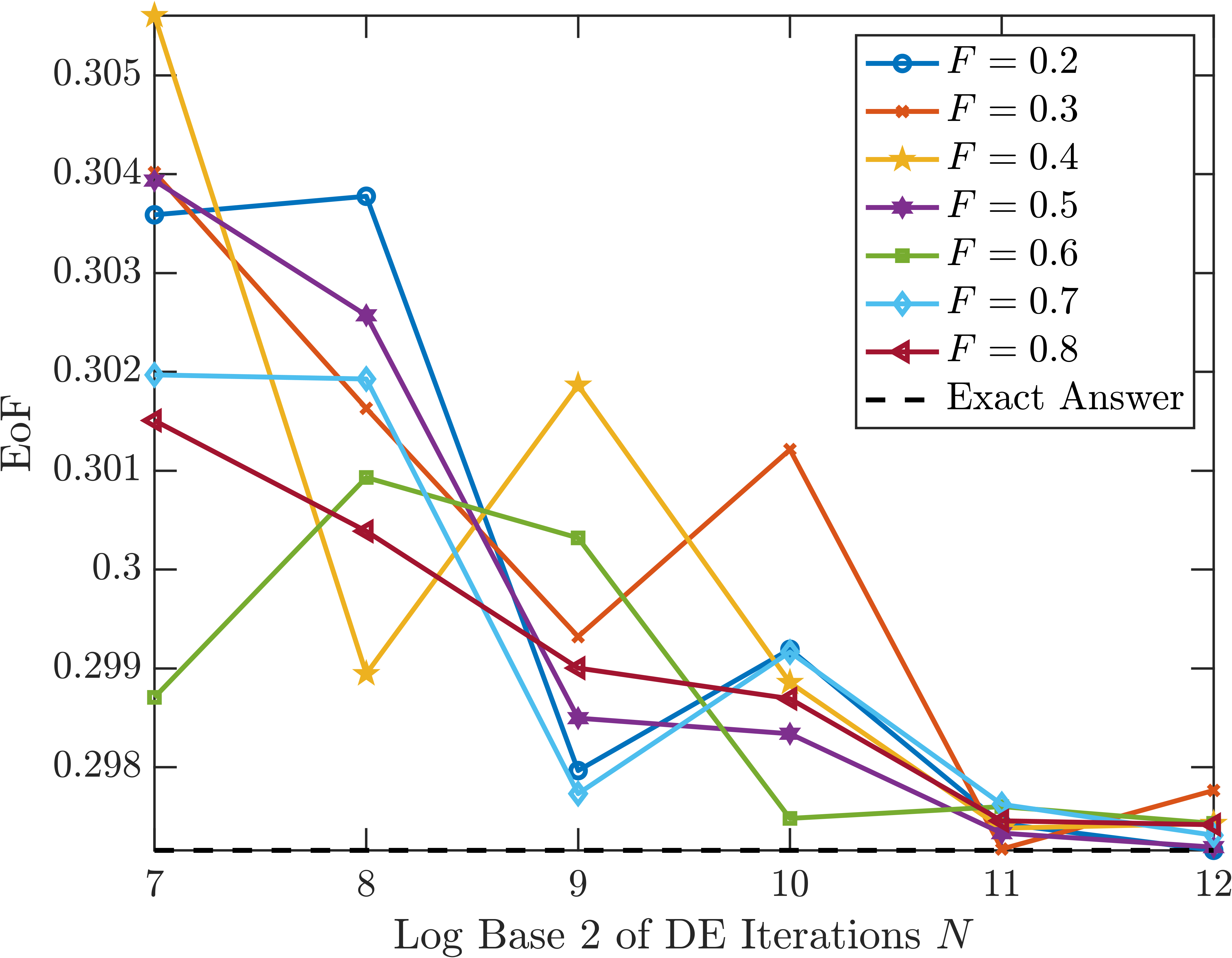}}
\subfigure{\includegraphics[width=0.45\textwidth]{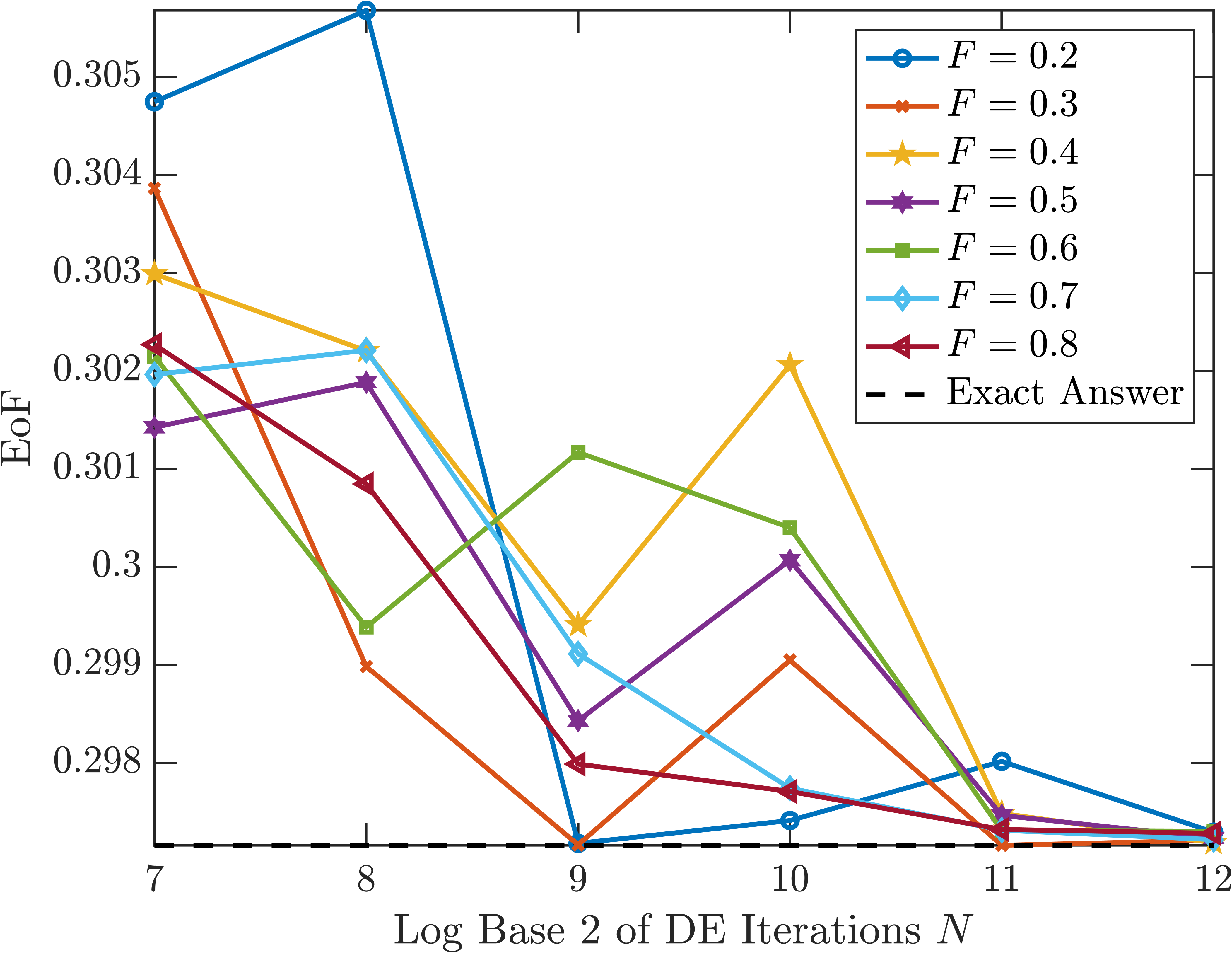}}
\subfigure{\includegraphics[width=0.45\textwidth]{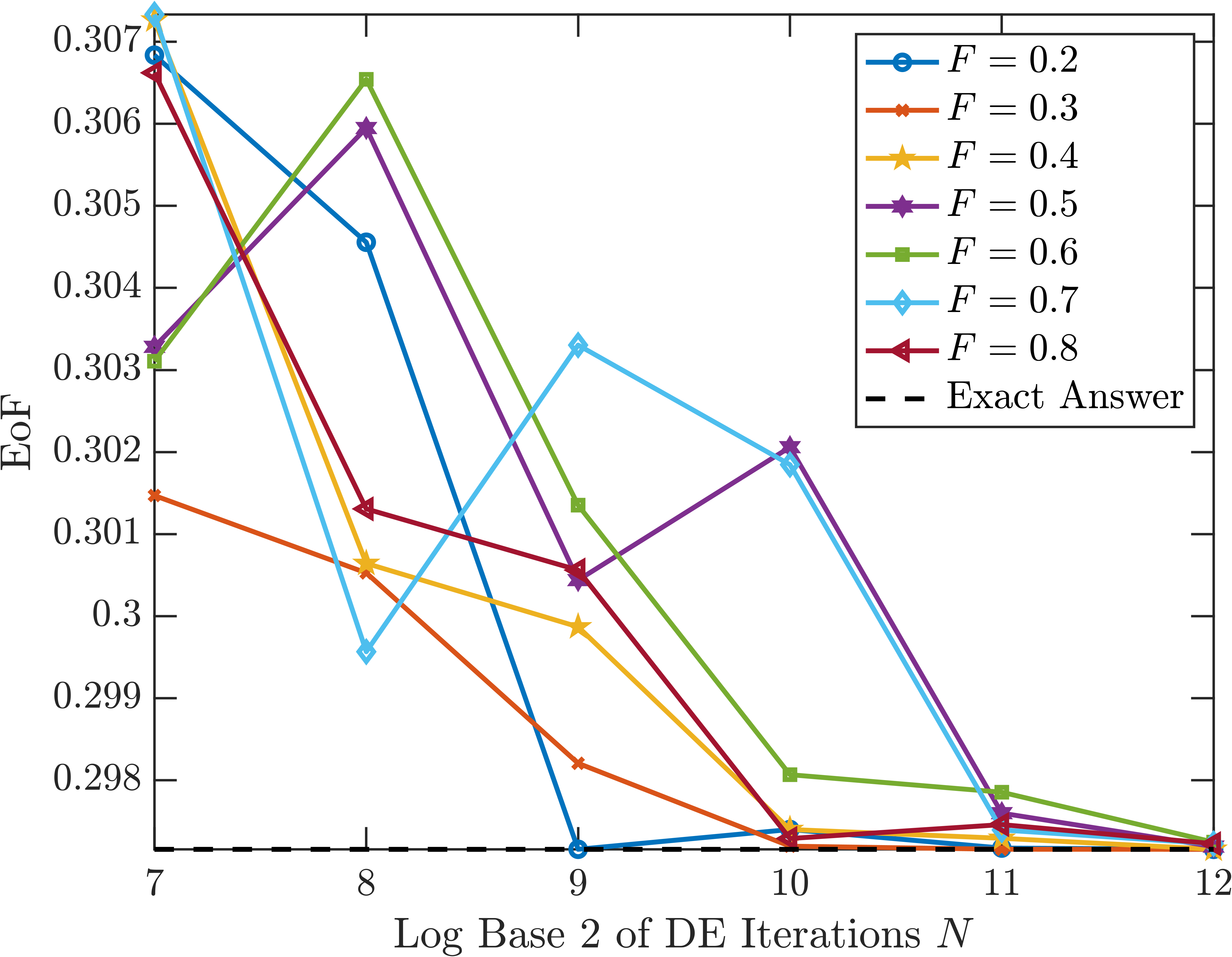}}
\end{centering}
\caption{The impact of varying DE parameters with $N_{\rm pop}$ fixed to 30. In each panel, we show the value of the objective function $J$ after an optimization has been performed on the density matrix $\rho_1$ from Equation~\eqref{eq:rho1}. The top left panel has $C_R=0.3$, the top right has $C_R=0.5$, the bottom left has $C_R=0.7$, and the bottom right has $C_R=0.9.$ We observe that over the many possible choices in the weighting $F$ and the crossover $C_R,$ a small weight and high crossover perform well. The optimal choice in this hyperparameter study, $F=0.1$ and $C_R=0.9$ is shown in the bottom right panel.}\label{fig:FirstExample}
\end{figure*}
 
\section{Computational Examples\label{sec4}}
As a proof of concept, we test our method on four examples of density matrices where the computation of the objective function $J$ in Equation~\eqref{eq:EasierOpt} is done with the entropy of entanglement given by Equation~\eqref{eq:EoE}; in this case, $J$ is referred to as the EoF. Although we only report the EoF in our numerical results, it is clear that our methodology can be used for any other entropy of entanglement.

Also note that DE requires a choice of parameters $F,\ C_R,$ and $N_{\rm pop}.$ For this reason, we use the first example as a way to also tune the parameters in the DE algorithm. 

\subsection{First Example: Decohered Bell-like state}

A decohered two-qubit Bell-like state, which is confined to the $\{|01\rangle,|10\rangle\}$ subspace
of the full two-qubit Hilbert space $\{|ij\rangle\}$, is the simplest example of a mixed state
that can be entangled. It is given by the 
density matrix
\begin{equation}\label{eq:rho1}
\rho_1=\begin{pmatrix}
    0 & 0 & 0 & 0 \\
    0 & b & x & 0 \\
    0 & x^* & 1-b & 0\\
    0 & 0 & 0 & 0
\end{pmatrix},
\end{equation}
for which the convex-roof entanglement measure coincides with its von Neumann entropy. Indeed, the von Neumann entropy can be computed from $\rho_1$'s eigenvalues
\begin{eqnarray}
\nonumber
\lambda_1&=&\frac{1}{2}\left(1+\sqrt{4b^2-4b+4|x|^2+1}\right),\\
\nonumber
\lambda_2&=&\frac{1}{2}\left(1-\sqrt{4b^2-4b+4|x|^2+1}\right),\\
\nonumber
\lambda_3&=&\lambda_4=0.
\end{eqnarray}
When $b=x=1/3,$ the von Neumann entropy is thus calculated to be $-\lambda_1\log\lambda_1-\lambda_2\log\lambda_2= 0.381264053728103$ to 15 digits of precision. 

Running Algorithm~\ref{algo:HDE} to solve optimization problem~\eqref{eq:EasierOpt} for the density matrix $\rho_1$, we find that $M(U_*)$, with the underscore $*$ signifying numerical optimality, evaluates correctly to 14 digits. We find this result after $2^{13}$ iterations with DE parameters $F=0.1,$ $C_R=0.9,$ and $N_{\rm pop}$=30. On an average workstation, this computation takes about 100 seconds. With a more manageable $2^{10}$ iterations, the optimization finds the correct answer to 6 digits in about 12 seconds.

Choosing the DE parameters is done empirically. Our empirical finding is that $F=0.1$ and $C_R=0.9$ consistently work well together. In Fig.~\ref{fig:FirstExample}, we show how the optimization converges as the number of iterations increases for different choices of $F$ and $C_R$. Although our numerical exploration with this hyperparameter study could be done more thoroughly, we find that our result is sufficient in practice. In the future, we commit to using these parameters in numerical computations.

Two more comments are in order about our results. The first is that the convergence in Fig.~\ref{fig:FirstExample} is not necessarily monotonic with an increasing number of iterations. The reason is that the optimization method is inherently stochastic. The realizations that we show are fairly representative of what to expect when using our method. The second is about the choice of $k\geq r$ defining the feasible space $\mathcal{U}(k,r).$ We empirically find that $k=r$ works sufficiently well here and in most of the remaining examples. This is in opposition to what is reported in~\cite{rothlisberger2009numerical}, where the choice $k=r+4$ is made. We find that our results are reliably more optimal for this study when $k=r=2$ than any value of $k$ between $r$ and $r+8.$ As it turns out, this simplifies the methodology slightly as a byproduct, since we now work with square unitary matrices as opposed to rectangular ones.

\subsection{Second Example: Sudden death and rebirth of entanglement}

In this example, we still restrict the study to the case of two qubits, so the exact value of the EoF
can be found without minimization using the Wootters formula \cite{wootters98}.
We study the time evolution of the entanglement of an initial state that is separable and mixed,
while the entanglement is driven by an iteraction between the $|1\rangle$ states of each qubit.
The time-dependent density matrix is given by
\begin{align}\label{eq:rho2}
\rho_2(t)=\frac{1}{4}\begin{pmatrix}
    1 & c & c & c^2 e^{i\omega t} \\
    c & 1 & c^2 & ce^{i\omega t} \\
    c & c^2 & 1 & ce^{i\omega t}\\
    c^2e^{-i\omega t} & ce^{-i\omega t} & ce^{-i\omega t} & 1
\end{pmatrix},
\end{align}
where $\omega\in\mathbb{R}$ and $t\in[0,T],\ T>0.$ 

The parameter $c\in[0,1]$ corresponds to pure single-qubit dephasing channels that act on both qubits
\cite{nielsen00} on the two-qubit pure state $|\psi\rangle=|\!\!+\!+\rangle$, with $|+\rangle=(|0\rangle
+|1\rangle)/\sqrt{2}$. The evolution is driven by an entangling interaction Hamiltonian
$H_{\mathrm{int}}=\hbar\omega|11\rangle\langle 11|$. The evolution of entanglement for such states
is known to display sudden death and sudden birth of entanglement for $c<1$ \cite{roszak06b}.

\begin{figure}[!tb]
\begin{centering}
\subfigure{\includegraphics[width=0.45\textwidth]{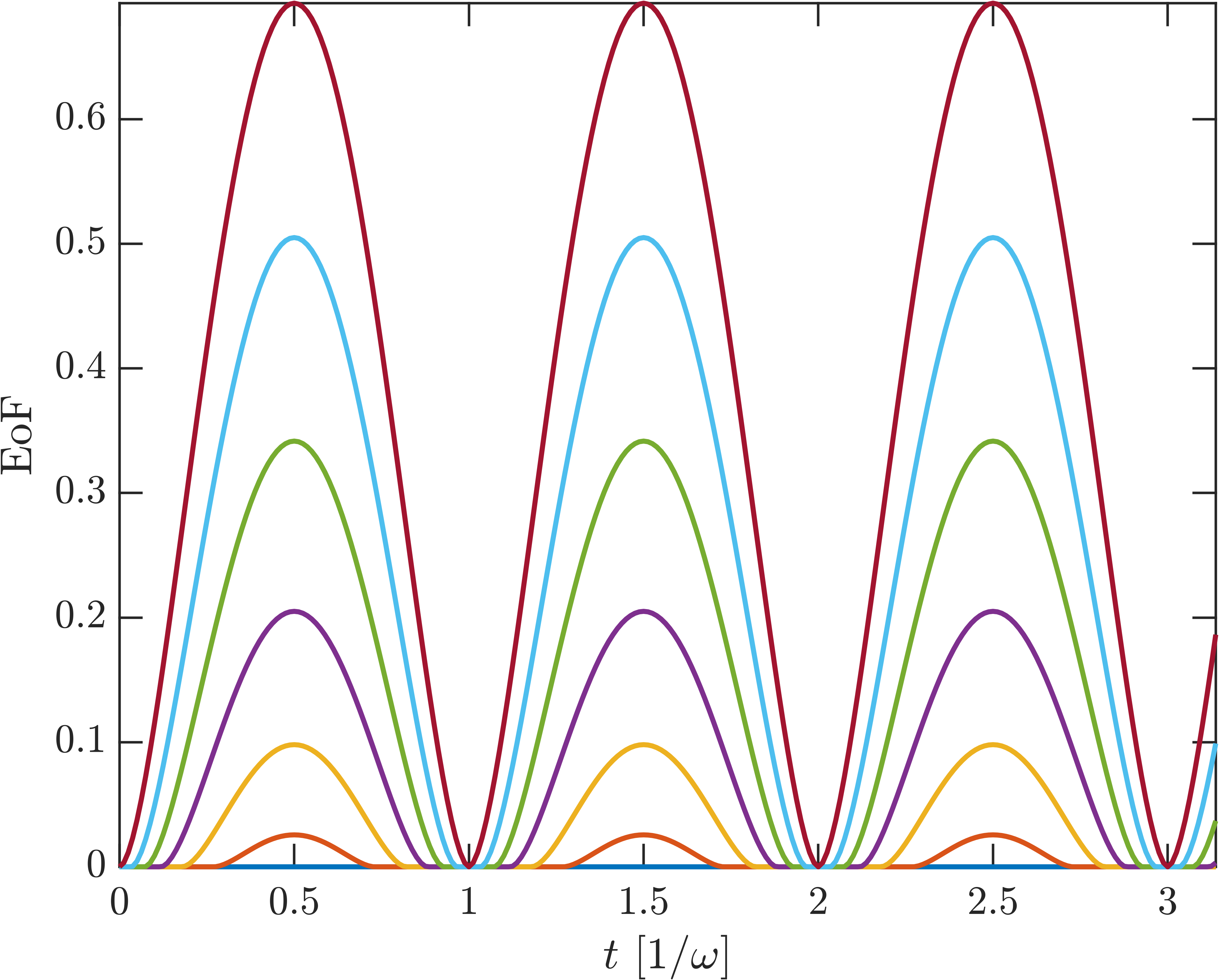}}
\end{centering}
\caption{
Evolution of two-qubit entanglement in state (\ref{eq:rho2}), for different values of the coherence parameter
$c$. The top curve corresponds to pure state evolution $c=1$, which cannot and does not display life/death cycles and reaches the maximum value of unnormalized EoF, $2\ln 2$.
Other curves correspond to mixed states and life/death cycles are observed.
The coherence parameter changes from top to bottom as follows, $c=1,0.9,0.8,0.7,0.6,0.5,0.4.$}\label{fig:lifedeath}
\end{figure}

In Fig.~\ref{fig:lifedeath}, we show the evolution of EoF obtained using our minimization method
for different values of the dephasing parameter $c$.
We find that not only the continuous, smooth evolution of entanglement is reproduced, but the life-death cycles for the mixed initial state
are very clearly visible 
(both in agreement with the Wootters formula). Thus, the method is capable not only of quantitative description of entanglement
present in a system, but can reliably reproduce evolution, including nontrivial features characteristic
for entanglement, such as sudden death. 

\subsection{Third Example: Qubit-environment entanglement evolution} 

\begin{figure*}[htbp]
\begin{centering}
\subfigure{\includegraphics[width=0.45\textwidth]{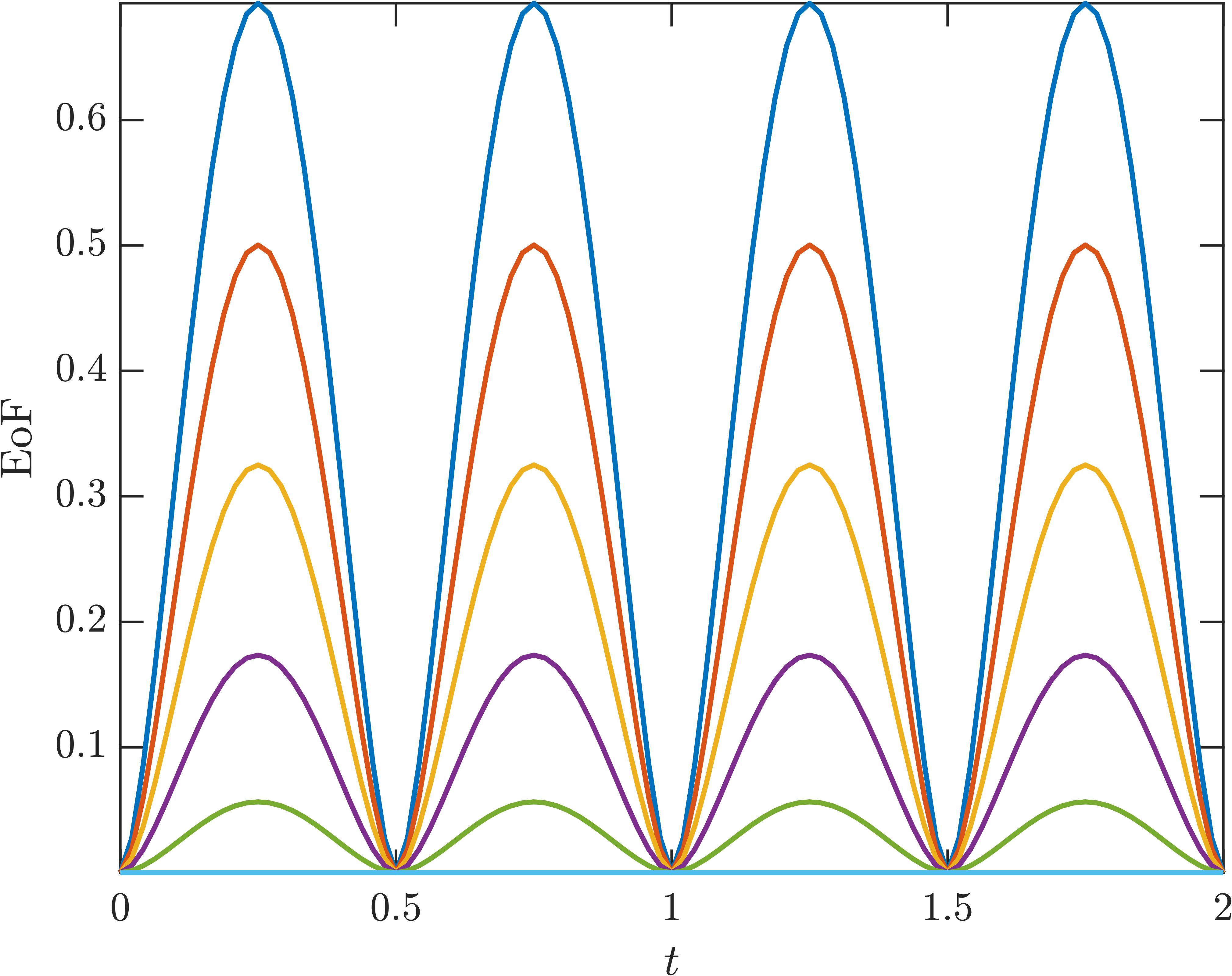}}
\subfigure{\includegraphics[width=0.45\textwidth]{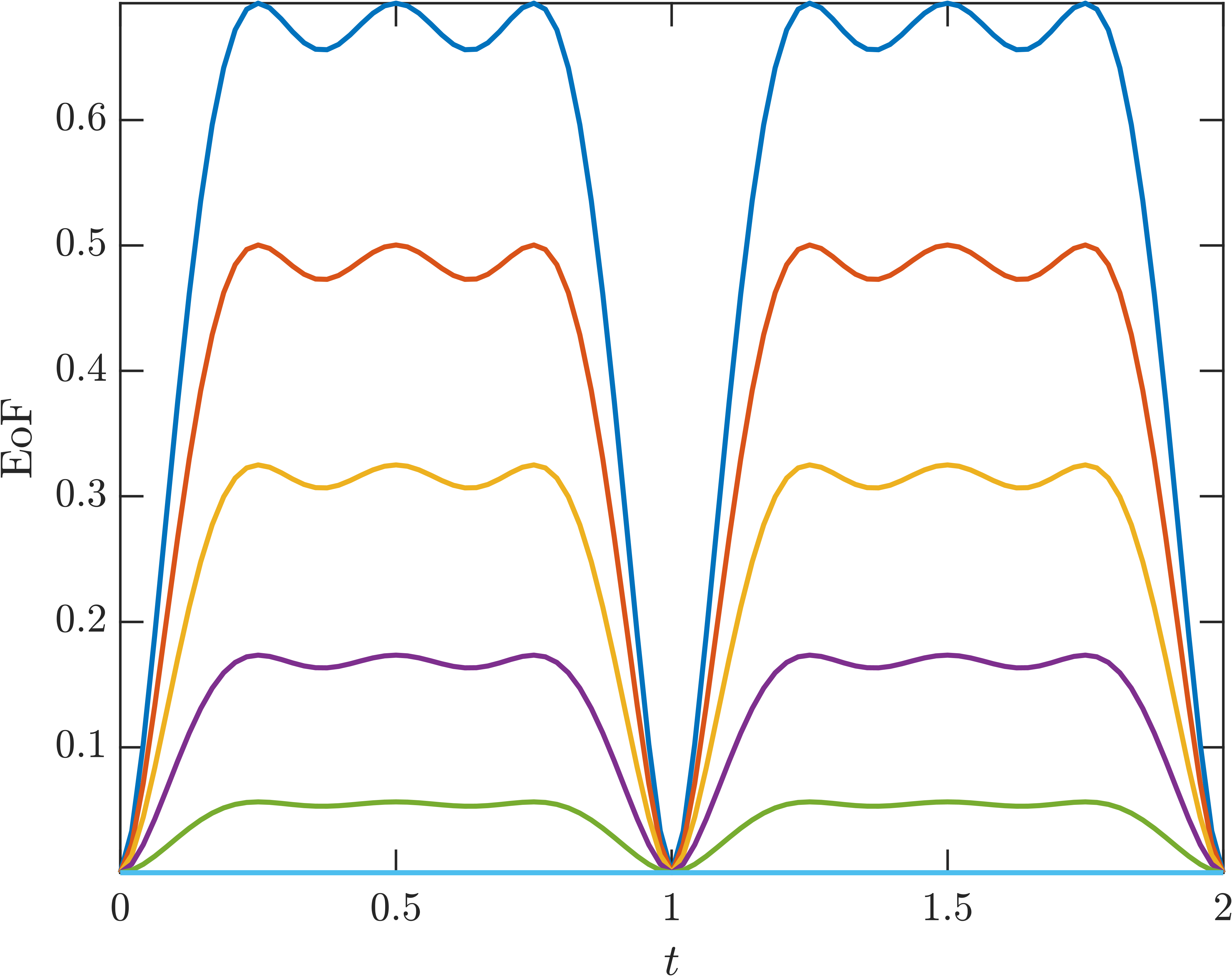}}
\subfigure{\includegraphics[width=0.45\textwidth]{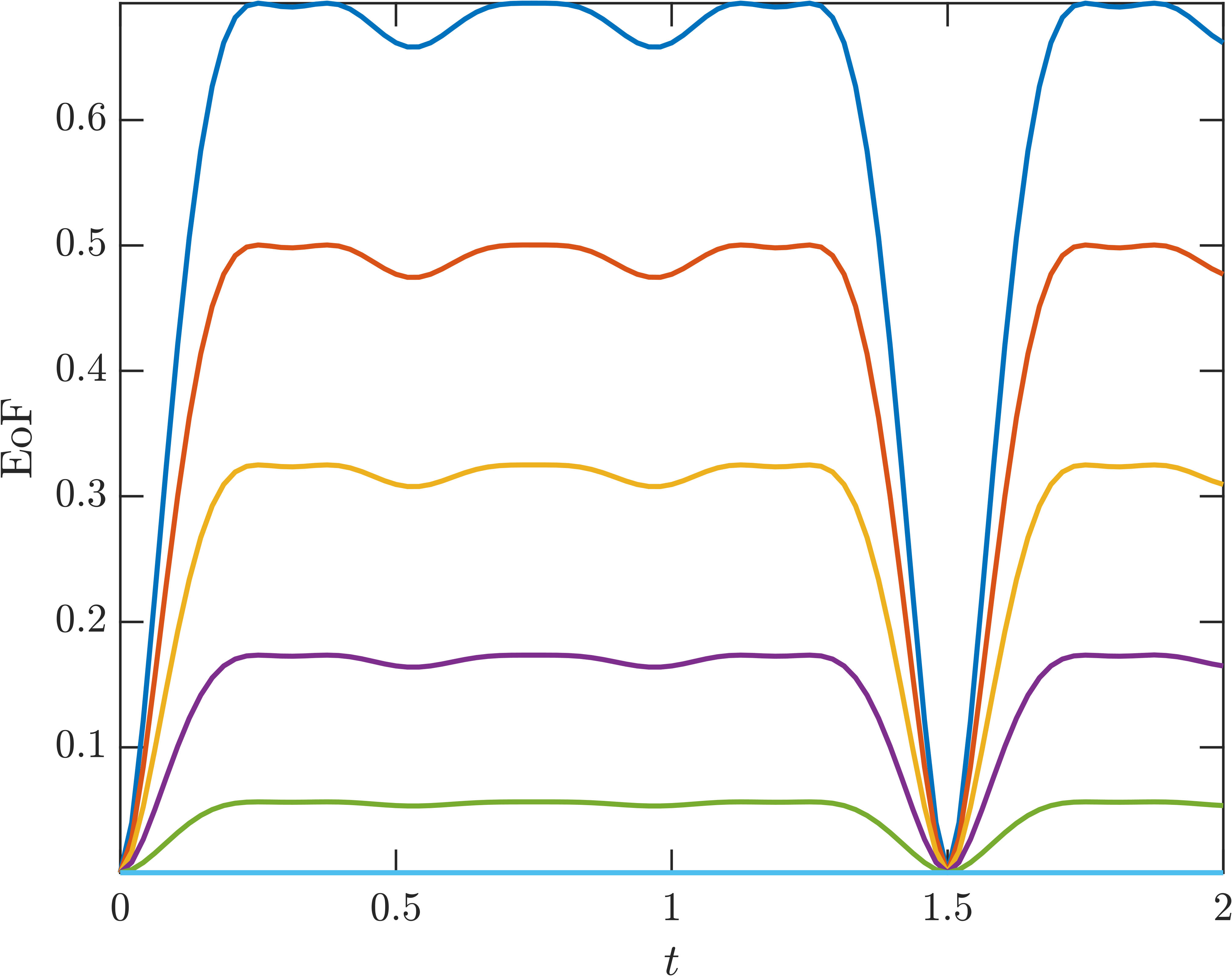}}
\subfigure{\includegraphics[width=0.45\textwidth]{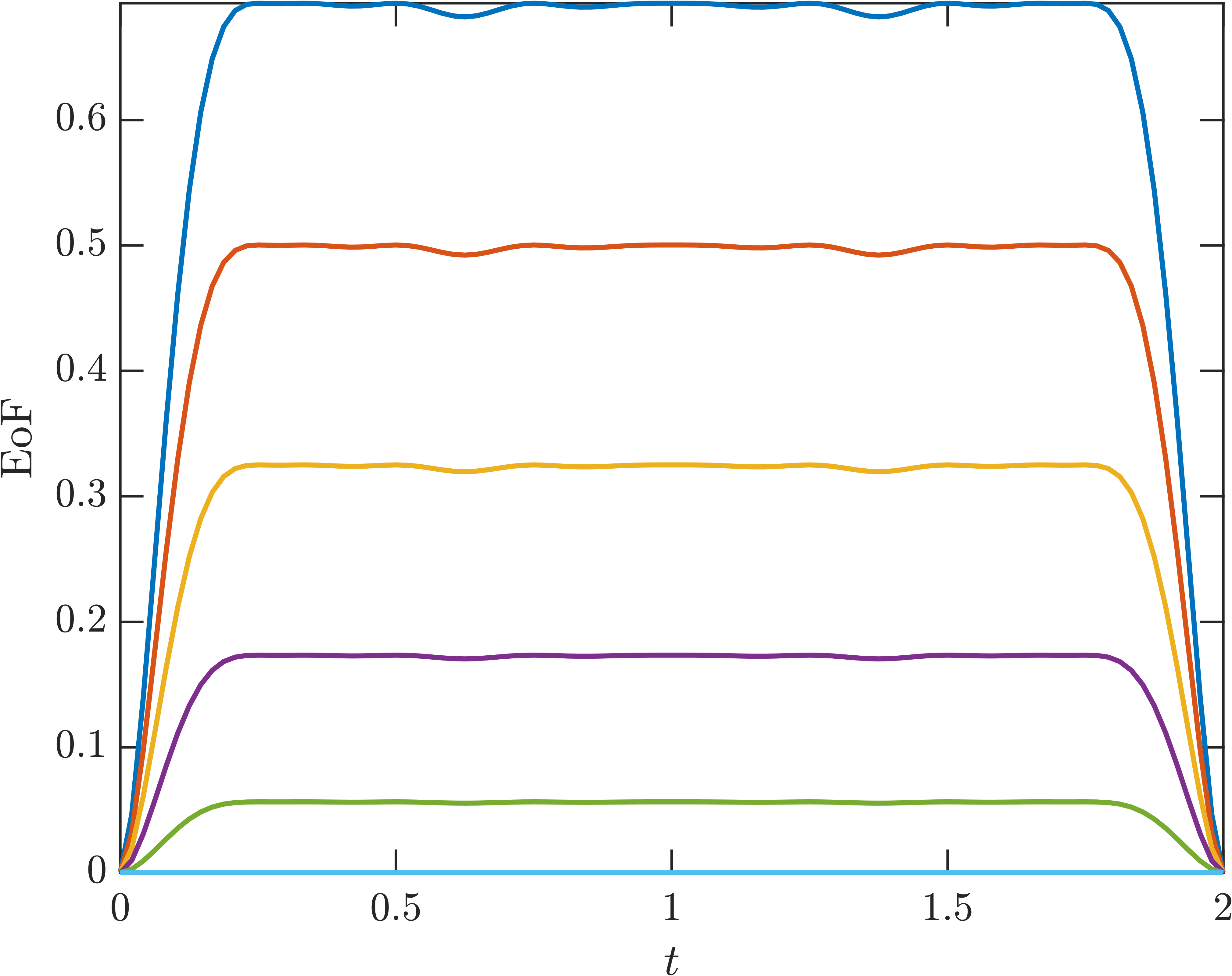}}
\end{centering}
\caption{Evolution of entanglement for density matrix (\ref{rhos}) for an environment of (top left) 2, (top right) 4, (bottom left) 8, and (bottom right) 16 qubits
for different purity of the initial qubit state, with $d=0, 0.2, 0.4, 0.6, 0.8$ and $1$.}\label{fig:CoT}
\end{figure*}

Since the algorithm parameters are set and we have demonstrated that it is capable of reproducing
continuous, smooth evolution, as well as sudden death of entanglement, in agreement with the two-qubit formula, which does not require minimization, we now turn to the study of larger systems.
In this case, there is no formula for EoF that does not require minimization, but there exist examples
of larger systems, for which the calculation of some entanglement measure is relatively straightforward.

One such example involves a qubit interacting with an environment of arbitrary size via an interaction
that can only lead to pure dephasing of the qubit. 
We study entanglement between a qubit and its environment composed of $N_{\rm e}$ environmental
qubits, following the example of Ref.~\cite{roszak20}. Contrarily to the results presented in that paper,
we will keep the environment pure, while making the initial qubit state mixed, producing a different set of curves.

The qubit is initially in a product state between the qubit and each environmental qubit.
The qubit is in a mixed state given by
\begin{equation*}
\rho_{\rm q}(0)=
\frac{1}{2}
\left(
\begin{matrix}
1&d\\
d&1
\end{matrix}
\right),
\end{equation*}
which corresponds to an equal superposition state that already underwent some pure dephasing, quantified by the real
parameter $d$.
Each environmental qubit (labelled by $k$) is initially in its ground state
$|0\rangle_k$.
Thus the density matrix of the whole environment is
\begin{equation}
\rho_{\rm e}(0)=\bigotimes_{k=1}^{N_{\rm e}} |0\rangle_{kk}\langle 0 |.
\end{equation}
The total density matrix of the system is given by the product state
\begin{equation}
    \rho_{\rm s}(0)=\rho_{\rm q}(0)\otimes\rho_{\rm e}(0).
\end{equation}

For pure-dephasing Hamiltonians, it is always possible to write the qubit-environment 
Hamiltonian in the form \cite{roszak18}
\begin{equation}
    H=\sum_{i=0,1}|i\rangle\langle i|\otimes V_i,
\end{equation}
where the operators $V_i$ act on the space of the environment.
Then the evolution operator for the whole system can be written as 
\begin{equation}
U(t)=\sum_{i=0,1}|i\rangle\langle i|\otimes w_i(t),
\end{equation}
with $w_i(t)=\exp(-iV_it/\hbar)$. The interpretation of such a process is that the evolution of the
environment is conditional on the qubit pointer state, and thus the source of the decoherence
is the distinguishability of the qubit state via the state of the environment.

In the following, we omit the free qubit evolution (which has no bearing on entanglement)
and the free evolution of the environment, while assuming that no correlations between different
environmental qubits is formed during the evolution. 
For simplicity, we also assume that the interaction is fully asymmetric, which means that
$w_0(t)=\mathbb{I}$, which does not have a meaningful bearing on the evolution of the entanglement
\cite{roszak15}.
Hence, the relevant environmental evolution operator is given by
\begin{equation}
\hat{w}_1(t)=\bigotimes_{k=1}^{N_{\rm e}} \hat{w}^k(t).
\end{equation}
Following Ref.~\cite{roszak20}, we take the evolution operators corresponding to each environmental
qubit to be of the form,
\begin{equation}
\hat{w}^k(t)=e^{i \omega_k t}\left|+\rangle\langle+\right|\ +\ e^{-i \omega_k t}\left|-\rangle\langle-\right|,
\end{equation}
with the states $| \pm\rangle=\frac{1}{\sqrt{2}}\left(|0\rangle \pm|1\rangle\right)$ and differing phases $\omega_k=\frac{2\pi}{k},$ $k=1,\cdots,N_{\rm e}$. 

The time-dependence of entanglement of the qubit-environment density matrix
\begin{equation}
\label{rhos}
    \rho_{\rm s}(t)=\hat{U}(t)\rho_{\rm s}(0)\hat{U}^{\dagger}(t).
\end{equation} 
is shown in Fig.~\ref{fig:CoT}.
The plotted evolution curves are for environments of two, four, eight, and sixteen qubits, the last of which corresponds to a Hilbert space of dimension
$2^{17}$ for the whole system; thus, the minimization is over an extremely large number of parameters. 
There is no other way to obtain the EoF curves other than minimization, but the results for the evolution of the entanglement
presented in Ref.~\cite{roszak20} allow a qualitative assessment of their validity.
The EoF curves we find are smooth and qualitatively in good agreement with the results of~\cite{roszak20}
where applicable. Moreover, the computation time for a 16 qubit environment is still a manageable 20 minutes for each moment along the time axis. We used parallel computing (24 cores) to maintain the computation of each entanglement curve over 96 moments in time to just under an hour and a half.

\subsection{Fourth 
Example: Temperature dependence of qubit-environment entanglement at equilibrium.} 

\begin{figure}[!tb]
\begin{centering}
\subfigure{\includegraphics[width=0.45\textwidth]{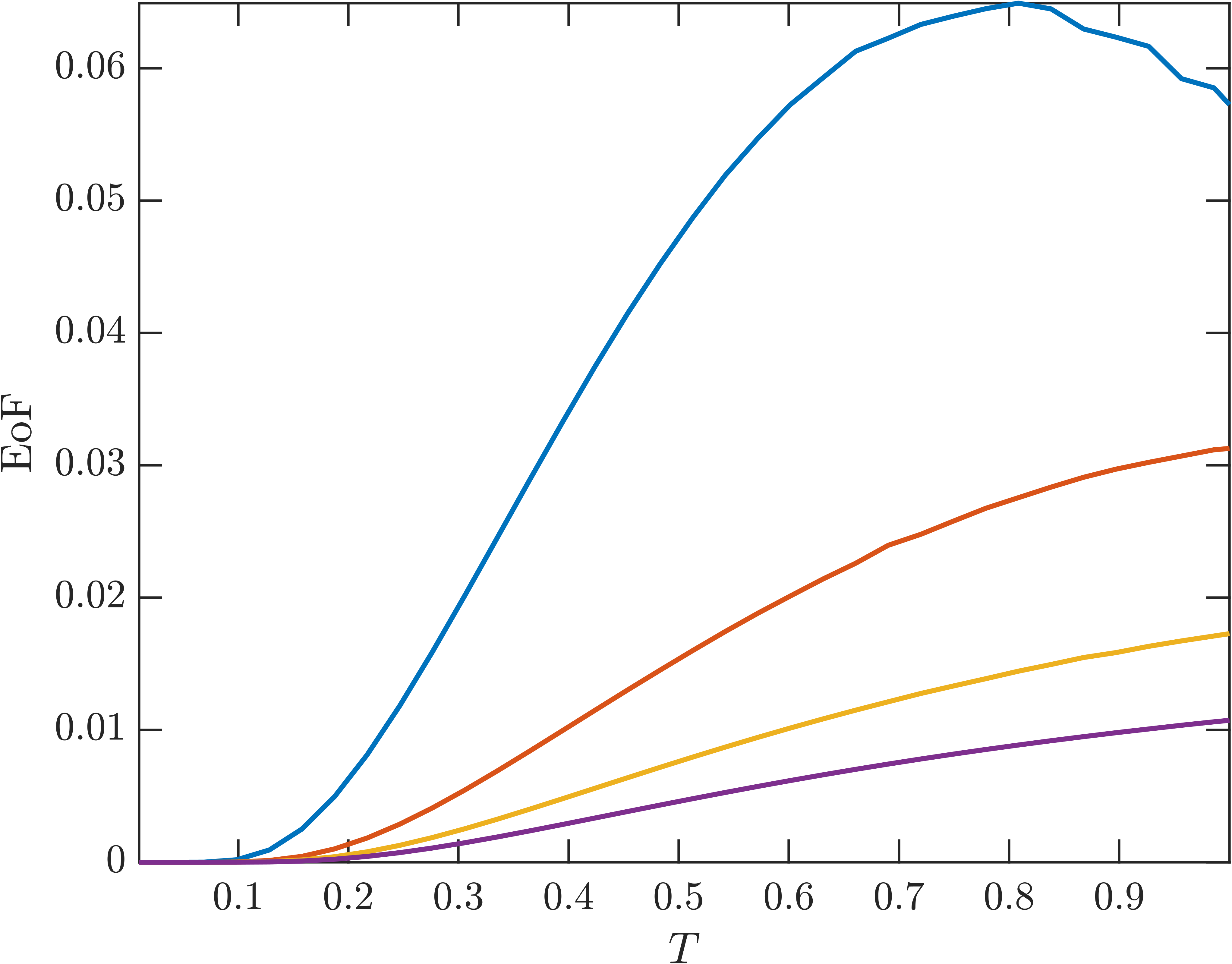}}
\subfigure{\includegraphics[width=0.45\textwidth]{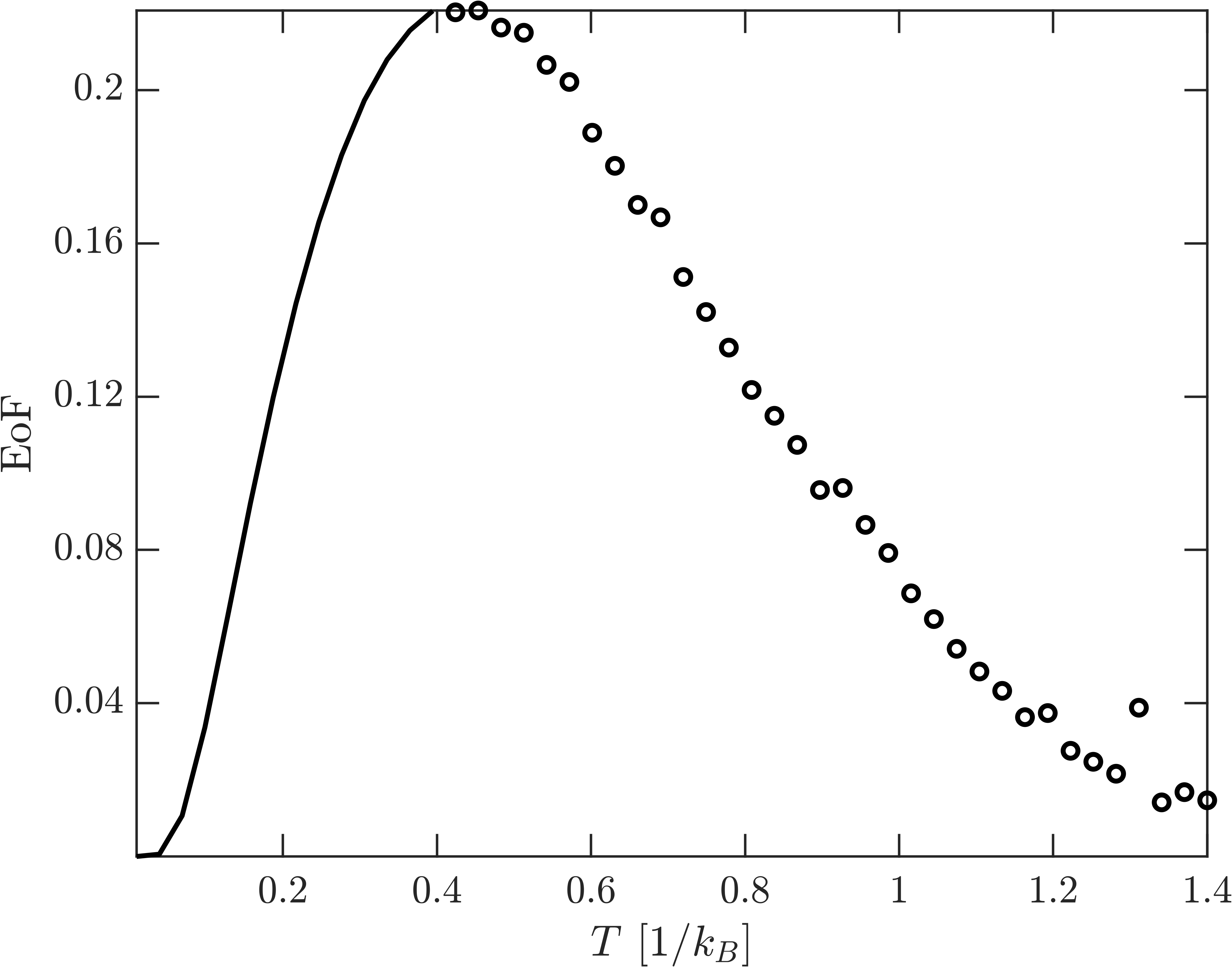}}
\end{centering}
\caption{
Temperature dependence of qubit-environment entanglement at equilibrium for a system with three blocks, $K=1$.
The top panel shows entanglement curves for different omega $\Omega$ (5, 10, 15 and 20 from bottom to top), which corresponds to different applied magnetic fields
in a spin system. The bottom panel shows a computation for a wider range of temperatures for $\Omega=1$.
We use points at higher temperatures, when the algorithm no longer produces a smooth temperature dependence.}\label{fig:Gibbs}
\end{figure}

Another example of a larger system for which entanglement can be efficiently computed is a density matrix that is block diagonal in disjoint
subspaces \cite{jedrzejewski22}. Following the example of Ref.~\cite{jedrzejewski22}, we 
use our algorithm to find the EoF of a thermal equilibrium state corresponding to a Hamiltonian that
has this special block form. 
Thus the density matrices under study are obtained from the Gibbs state,
\begin{equation}
\label{gibbs}
    \rho_{\rm Gibbs}=\frac{e^{-H/k_BT}}{\mathrm {Tr} \ e^{-H/k_BT}}.
\end{equation}
The Hamiltonian that enters eq.~(\ref{gibbs}) is a qubit-environment Hamiltonian and
is given by $H=\sum_m H_m$, where $m$ numbers the blocks.
Each block is given by \cite{jedrzejewski22}
\begin{equation}
\label{hm}
H_m=\left(\begin{array}{cccc}
E_{m_1} & 0 & 0 & M_m \\
0 & E_m & 0 & 0 \\
0 & 0 & E_m & 0 \\
M_m^* & 0 & 0 & E_{m_2}
\end{array}\right).
\end{equation}
The critical difference between the blocks is that the subspace of each block is disjoint.
By this we mean that in the effectively two-qubit subspace of each block, the two states corresponding to the
qubit remain unchanged, but the states corresponding to the environment are different in each block
(form a separate subspace).
Thus, the matrix for a single block of the Hamiltonian (\ref{hm}) is written in the bases' subspace
$\{|0m\rangle,|0m_{\perp}\rangle,|0m\rangle,|0m_{\perp}\rangle\}$,
where $\rangle m|m_{\perp}\rangle = 0$, but also all states that occupy different subspaces are orthogonal to each other.
This requirement is stronger than for the Hamiltonian to simply have block-diagonal form, but it allows for
convex roof entanglement measures to be averaged over different blocks without the necessity of minimization beyond
each block \cite{jedrzejewski22}. This gives us the opportunity to compare the results obtained by full minimization
performed by our algorithm with the results of~\cite{jedrzejewski22}, which took advantage of the special feature.

The parameters within each block of the Hamiltonian (\ref{hm}) are given by
\begin{subequations}
\begin{eqnarray}
E_{m_1} & =&\alpha\left(m+\frac{\Omega}{2}\right), \\
E_{m_2} & =&-\alpha\left(m+1+\frac{\Omega}{2}\right), \\
E_m&=&\frac{1}{2}\left(E_{m_1}+E_{m_2}\right),\\
M_m & =&\alpha \sqrt{K(K+1)-m(m+1)},
\end{eqnarray}
\end{subequations}
where $m=0, \pm 1, \pm 2, \ldots, \pm K$.
The parameter $\alpha$ (which we set to $1$) dictates the coupling strength, while
$\Omega$ is responsible for the energy splittings within the system.

The temperature dependence of the equilibrium entanglement between a qubit and an environment with $K=1$, which corresponds
to a Hamiltonian with three blocks, is shown in Fig.~\ref{fig:Gibbs}. 
Note that a three-block environment translates to six environmental states, so we use the algorithm to quantify
entanglement within a $12$-dimensional Hilbert space.

For larger values of $\Omega$ (top panel), which roughly correspond
to large magnetic fields in spin systems, the resulting curves are smooth and in agreement with corresponding 
results of Ref.~(\cite{jedrzejewski22}) up to reasonably high temperatures (but not for the whole scope of non-negligible entanglement). The EoF as a function of the temperature of $\Omega=1$, is plotted for a wider
range of temperatures in the bottom panel. At low temperatures, the curve obtained is smooth and does not differ qualitatively from the higher$\Omega$ curves. At larger temperatures
which exceed the temperature at which maximum entanglement is reached, convergence cannot be reached in a
computation time of a whole day. Despite this, the general trend is still visible and it agrees very well with the results of Ref.~\cite{jedrzejewski22}.

\section{Conclusion\label{sec5}}

We developed a method for the efficient computation of mixed-state convex roof entanglement measures.
Such measures require minimization over all pure-state decompositions of the density matrix.
Operationally, the task requires minimization over a set of unitary matrices
of the same dimension as the Hilbert space of the problem under study, $N$, and the number of minimization parameters
scales as $N^2-1$. Our method thoroughly searches this space using a nonconvex method that remains constrained to the manifold of unitary matrices via a QR factorization. The convergence to the nearest local minimum is then accelerated by using a BFGS method.

We used the developed algorithm on a number of examples. The examples were chosen in such a way that at least the 
trends of entanglement behavior would be known. Furthermore, the studied scenarios involved smooth or partially smooth
parameter-dependence (time or temperature) of entanglement in order to provide a better test of the reproducibility 
and reliability of the outcomes. We have found that the algorithm is extremely reliable (and fast) in identifying the expected trends as long as the
overall purity of the state is fairly high. This holds true for very large systems
(in the third example we produce entanglement evolution curves for a Hilbert state of dimension $2^{17}$).

The algorithm is unable to quantify the EoF with high precision at low purity, as seen at high temperatures in the fourth
example. This offers two important opportunities for future work. The first is mathematical and is to explore why the algorithm is unable to thoroughly search the non-convex landscape at low purities. We conjecture that this is due to an increasing malignant nonconvexity as the purity decreases and thus makes the identification of the global optimum unlikely. 

The second direction is computational. As can be gleaned from the presentation of our methodology, any black-box numerical optimization method can be used, in principle, if the QR factorization is deployed within the objective function evaluation. This offers an opportunity to explore the numerical optimization methods typically used by the machine learning community, such as adaptive moment estimation~\cite{kingma2014adam}, that address high-dimensional nonconvex optimization problems.

\section{Acknowledgments}
We appreciate the help of Alejandro Aceves in preparing this manuscript for publication. J.A. also acknowledges support from NSF award number 2316622.
K.R.: This project is funded within the QuantERA II Programme that has received funding from the EU H2020 research and innovation programme under GA No 101017733, and with funding organisation MEYS.

\end{document}